\documentclass{aa}
\usepackage{graphicx}
\usepackage{txfonts}
\newcommand{\RSTAR}{\mbox{$R_{\star}$}}

\newcommand{\MSOLPERYR}{\mbox{$M_{\sun}$~yr$^{-1}$}}
\newcommand{\micron}{\mbox{$\mu$m}}
\newcommand{\KMS}{\mbox{km s$^{-1}$}}
\newcommand{\HOH}{\mbox{H$_2$O}}
\newcommand{\PERSQCM}{\mbox{cm$^{-2}$}}
\newcommand{\HRK}{\mbox{HR\_K}}
\newcommand{\MRK}{\mbox{MR\_K}}
\newcommand{\NCOOUT}{\mbox{$N_{\rm out}$}}
\newcommand{\TIN}{\mbox{$T_{\rm in}$}}

\newcommand{\VFLOW}{\mbox{$v_{\rm flow}$}}
\begin{document}
\title{
Spatially resolving the inhomogeneous structure of the dynamical atmosphere 
of Betelgeuse with VLTI/AMBER
\thanks{
Based on AMBER and VINCI observations made with the Very Large Telescope 
Interferometer of 
the European Southern Observatory. Program IDs: 080.D-0236 (AMBER Guaranteed 
Time Observation), 60.A-9054A, and 60.A-9222A.}
}

\author{K.~Ohnaka\inst{1} 
\and
K.-H.~Hofmann\inst{1} 
\and
M.~Benisty\inst{2}
\and
A.~Chelli\inst{3}
\and
T.~Driebe\inst{1} 
\and
F.~Millour\inst{1,4}
\and
R.~Petrov\inst{4}
\and
D.~Schertl\inst{1}
\and
Ph.~Stee\inst{5}
\and
F.~Vakili\inst{4}
\and
G.~Weigelt\inst{1} 
}

\offprints{K.~Ohnaka}

\institute{
Max-Planck-Institut f\"{u}r Radioastronomie, 
Auf dem H\"{u}gel 69, 53121 Bonn, Germany\\
\email{kohnaka@mpifr-bonn.mpg.de}
\and
INAF-Osservatorio Astrofisico di Arcetri, Instituto Nazionale di 
Astrofisica, Largo E. Fermi 5, 50125 Firenze, Italy
\and
Laboratoire d'Astrophysique de Grenoble, UMR 5571, 
Universit\'e Joseph Fourier/CNRS, BP 53, 38041 Grenoble Cedex 9, France
\and
Lab. H. Fizeau, CNRS UMR 6525, Univ. de Nice-Sophia Antipolis, 
Obs. de la C\^{o}te d'Azur, Parc Valrose, 06108 Nice, France
\and
Lab. H. Fizeau, CNRS UMR 6525, Univ. de Nice-Sophia Antipolis, 
Obs. de la C\^{o}te d'Azur, Avenue Copernic, 06130 Grasse, France
}

\date{Received / Accepted }

\abstract
{
}
{
We present spatially resolved high-spectral resolution $K$-band 
observations of the red supergiant Betelgeuse 
($\alpha$~Ori) using AMBER at the Very Large Telescope Interferometer (VLTI).  
Our aim is to probe inhomogeneous structures in the dynamical atmosphere 
of Betelgeuse. 
}
{
Betelgeuse was observed in the wavelength range between 2.28 and 
2.31~\micron\ with VLTI/AMBER using 
baselines of 16, 32, and 48~m.  The spectral 
resolutions of 4800--12000 allow us to study 
inhomogeneities seen in the individual CO first overtone lines.  
}
{
Spectrally dispersed interferograms have been successfully obtained 
in the second, third, and fifth lobes, which 
represents the highest spatial resolution (9~mas) achieved 
for Betelgeuse.  This corresponds to five resolution elements over its 
stellar disk.  
The AMBER visibilities and closure phases in the $K$-band continuum 
can be reasonably fitted by a uniform disk 
with a diameter of $43.19 \pm 0.03$~mas or a limb-darkening disk with 
$43.56 \pm 0.06$~mas and a limb-darkening parameter of 
$(1.2 \pm 0.07) \times 10^{-1}$. 
These AMBER data and the previous $K$-band 
interferometric data taken 
at various epochs suggest that Betelgeuse seen in the $K$-band continuum 
shows much smaller deviations from the above uniform disk or limb-darkened 
disk than predicted by recent 3-D convection simulations for red supergiants. 
On the other hand, our AMBER data in the CO lines reveal 
salient inhomogeneous structures.  
The visibilities and phases (closure phases as well as differential phases 
representing asymmetry in lines with respect to the continuum) 
measured within the CO lines 
show that the blue and red wings originate in spatially 
distinct regions over the stellar disk, indicating an inhomogeneous 
velocity field which makes the star appear different in the blue and red 
wings.  
Our AMBER data in the CO lines can be roughly explained by a simple 
model, in which a patch of CO gas is moving outward or inward at velocities 
of 10--15~\KMS, while the CO gas in the remaining region in the atmosphere 
is moving in the opposite direction at the same velocities.  
Also, the AMBER data are consistent with the presence 
of warm molecular layers (so-called MOLsphere) 
extending to $\sim$1.4--1.5~\RSTAR\ with a CO column density 
of $\sim \! 1 \times 10^{20}$~\PERSQCM. 
}
{
Our AMBER observations of Betelgeuse are the 
first spatially resolved study of the 
so-called macroturbulence in a stellar atmosphere (photosphere and possibly 
MOLsphere as well) other than the Sun.  
The spatially resolved CO gas motion is likely to be related to convective motion 
in the upper atmosphere or intermittent mass ejections in clumps or arcs.  
}


\titlerunning{Spatially resolving the inhomogeneous 
atmosphere of Betelgeuse}
\authorrunning{Ohnaka et al.}
\maketitle

\section{Introduction}
\label{sect_intro}

Red supergiants (RSGs) experience slow, intensive mass loss up to 
$10^{-4}$~\MSOLPERYR. 
Despite its importance not only in stellar evolution but also 
in the chemical enrichment of the interstellar matter, 
the mass loss mechanism in RSGs is not well understood.  
While radiation pressure on dust grains is often considered to be the 
driving mechanism of mass loss in cool evolved stars, 
it is not clear where and how dust forms in RSGs 
and how mass outflows are initiated. 
Alternative scenarios include Alfv\'en-wave-driven winds 
(Airapetian et al. \cite{airapetian00}; 
Schr\"oder \& Cuntz \cite{schroeder05}, \cite{schroeder07}), 
combination of Alfv\'en waves and the wave damping due to dust 
(Vidotto \& Janteco-Pereira \cite{vidotto06}), and 
convective turbulence combined with radiation pressure on 
molecules (Josselin \& Plez \cite{josselin07}).

The atmosphere of RSGs exhibits complicated structures.  In the lower 
photosphere, vigorous convective motion is expected with the convective cell 
size possibly comparable to the stellar radius (Schwarzschild 
\cite{schwarzschild75}; Freytag et al. \cite{freytag02}).  
The photometric variabilities, as well as the variations in the radial 
velocities of the metal lines in the visible, can be interpreted as 
caused by such giant convective cells (Kiss et al. \cite{kiss06}; 
Gray \cite{gray08}).  
In the outer region, extended chromospheres exist.  
For example, the UV observations of the M supergiant Betelgeuse 
($\alpha$~Ori) with the Hubble Space
Telescope revealed that the hot ($\sim$6000--8000~K) chromospheric 
plasma is more than twice as extended as the photosphere measured in 
the near-IR with a bright feature 
(Gilliland \& Dupree \cite{gilliland96}; 
Uitenbroek et al. \cite{uitenbroek98}) .   
The image in the H$\alpha$ line is even more extended, 4--5 times as 
large as the photosphere (Hebden et al. \cite{hebden87}), consistent 
with the extended chromosphere.  
However, radio continuum observations of Betelgeuse with 
the Very Large Array (VLA) suggest that much cooler ($\sim$1000--3000~K) 
gas extends to several stellar radii, showing an irregular structure 
(Lim et al. \cite{lim98}).  
Non-spherical shape of the outer atmosphere of Betelgeuse was also 
detected by mid-IR interferometric observations by 
Tatebe et al. (\cite{tatebe07}).  
Furthermore, the narrow-slit spectroscopy of Betelgeuse 
in the 10~\micron\ region by Verhoelst et al. (\cite{verhoelst06}) 
revealed that silicate dust forms only at large distances from the star 
($\ga$20~\RSTAR) and that Al$_2$O$_3$ may form as close as $\sim$2~\RSTAR.   
This means that the hot chromospheric plasma, cooler gas, and Al$_2$O$_3$ dust 
may coexist within several stellar radii from the star, but 
the cooler component is much more abundant compared to the chromospheric 
gas, because it dominates the radio emission.

The presence of cool gas in the outer atmosphere of RSGs 
is consistent with dense molecular layers existing close to the star, 
the so-called ``MOLsphere'', which was proposed 
by Tsuji (\cite{tsuji00a}; \cite{tsuji00b}) to explain the 
IR spectra of the early M supergiants $\alpha$~Ori and $\mu$~Cep.  
While these stars were deemed to be too hot for \HOH\ to form, 
he showed that dense 
\HOH\ gas with column densities of the order of $10^{20}$~\PERSQCM\ 
and temperatures of 1500--2000~K at $\sim$1.3--2.0~\RSTAR\ can 
explain the spectral features at 2.7 and 6~\micron\ which cannot be 
reproduced by non-gray hydrostatic photospheric models.  
Near- and mid-IR interferometric studies also lend support to the 
presence of the MOLsphere toward RSGs (e.g., Perrin et al. \cite{perrin04}, 
\cite{perrin05}, \cite{perrin07}; Ohnaka \cite{ohnaka04a}; 
Tsuji \cite{tsuji06}), 
although the current, crude MOLsphere models cannot reproduce the 
\HOH\ absorption lines observed at 12~\micron\ 
(Ryde et al. \cite{ryde06a}; \cite{ryde06b}).  
On the other hand, Verhoelst et al. (\cite{verhoelst09}) argue 
against the presence of such dense molecular gas in the outer atmosphere of 
RSGs. 
They instead propose that the 2.7 and 6~\micron\ features in RSGs 
can be explained by a continuous (i.e. featureless) dust opacity 
source such as amorphous carbon and metallic iron, although it is not 
clear whether or not such grain species indeed form in oxygen-rich 
environments.  

A better understanding of the inhomogeneous structure of the outer 
atmosphere of RSGs is a key to unraveling the mass-loss mechanism 
in these stars.  
Inhomogeneities over the stellar surface were detected 
by high spatial resolution imaging of a few nearby RSGs. 
The high-resolution images of Betelgeuse at 0.7--1.25~\micron\ with spatial 
resolutions of down to 30~mas (the stellar angular size is $\sim$50~mas 
at these wavelengths) show the 
wavelength-dependent appearance of asymmetric structures 
(Burns et al. \cite{burns97}; Tuthill et al. \cite{tuthill97}; 
Young et al. \cite{young00}).  
However, their origin is by no means clear.  
They may be related to large convective cells predicted to be present 
in cool luminous stars (Schwarzschild \cite{schwarzschild75}; 
Freytag et al. \cite{freytag02}) 
or alternatively to thermal instability taking place 
in the outer atmosphere.  For example, the magnetohydrodynamical 
simulation of Suzuki (\cite{suzuki07}) for red giant branch (RGB) 
stars, which are much less luminous than RSGs, 
shows that thermal instability leads to ``structured'' stellar 
winds with many bubbles of hot gas ($\sim \! 10^5$~K) 
embedded in cool winds ($\sim$1--5$\times 10^3$~K).  
For cooler RSGs, the formation of molecules may also promote 
such thermal instability, and in particular, CO is an important 
coolant in the atmosphere of late-type stars 
(Cuntz \& Muchmore \cite{cuntz94}).  

To glean clues to the origin of the inhomogeneities and the mass-loss 
mechanism in RSGs, 
high spatial resolution observations in IR molecular lines 
are very effective. 
The high spectral resolution ($\lambda/\Delta \lambda$=12000) 
of the near-IR interferometric instrument AMBER 
(Astronomical Multi-BEam combineR) 
at VLTI allows us to resolve 
the CO first overtone lines and to spatially resolve inhomogeneous structures 
within each CO line.  
In this paper, we present high-spectral and high-spatial resolution 
$K$-band AMBER 
observations of the prototypical RSG Betelgeuse (M1-2Ia--Ibe).

\section{Observations}
\label{sect_obs}

\subsection{AMBER}
AMBER (Petrov et al. \cite{petrov07}) operates in the $J$, $H$, and $K$ bands 
with spectral resolutions 
of 35, 1500, and 12000, combining three 8.2~m Unit Telescopes (UTs) 
or 1.8~m Auxiliary Telescopes (ATs).  
AMBER is a spectro-interferometric instrument which records 
spectrally dispersed fringes on the detector.  
With the maximum baseline length of 130~m currently available at VLTI, 
spatially resolved spectroscopy with an angular resolution of down to 2~mas is 
possible with AMBER, which enables us to study 
the wavelength dependence of the size and shape of the object.  
AMBER observations with three telescopes 
allow us to measure three visibilities and three differential phases (DPs), 
as well as one closure phase (CP).  
Visibility is the amplitude of the Fourier transform (complex function) of the 
object's intensity distribution in the plane of the sky and contains 
information about the size and shape of the object.  
On the other hand, the phase of the Fourier transform (also called ``Fourier 
phase'' to avoid confusion) contains information 
about the object's deviation from point-symmetry.  
While the atmospheric turbulence prevents us from measuring directly the 
phase, AMBER measures two observables (DP and CP) which are related to the 
object's phase.  
DP approximately represents the object's phase in a spectral feature 
measured with respect to the continuum\footnote{Exactly speaking, two pieces 
of information are lost in the derivation of DP from AMBER observations: 
the absolute phase offset and the linear phase gradient with respect to 
wavenumber.}.  Therefore, non-zero DPs mean a photocenter shift of the 
spectral feature forming region with respect to the continuum forming 
region.  
CP is the sum of phases around a closed triangle of baselines 
(i.e., $\varphi_{12} + \varphi_{23} + \varphi_{31}$) and not affected by the 
atmospheric turbulence.  For point-symmetric objects, CP is always zero 
or $\pi$.  
Non-zero and non-$\pi$ CPs, whether in the 
continuum or in some spectral features, indicate asymmetry of the object. 
Two-telescope AMBER observations provide only one visibility 
and one DP (CP cannot be measured with two telescopes).

Betelgeuse was observed on 2008 January 08 with AMBER using three ATs 
in the E0-G0-H0 linear array configuration 
with 16--32--48~m baselines (AMBER Guaranteed Time Observation, 
Program ID: 080.D-0236A, P.I.: K.~Ohnaka).  
Since the E0-G0-H0 configuration is 
a linear array lying at $+$71\degr\ from North ($+$90\degr\ = East) on 
the ground, 
the position angles of the three projected baseline vectors are the same.  
We used the $K$-band high-resolution mode (HR\_K) of AMBER with a spectral 
resolution of 12000 covering wavelengths from 2.28 to 2.31~\micron.  
As shown below, this wavelength range was chosen to observe the strong 
$^{12}$C$^{16}$O (hereafter simply CO) first overtone lines near the (2,0) 
band head.  
The $H$-band brightness of Betelgeuse is too high for the VLTI fringe tracker 
FINITO.  However, the extremely high brightness of Betelgeuse 
($K$ = $-4.4$), 
together with the excellent weather conditions (0\farcs 3--0\farcs4 seeing), 
enabled us to detect low-contrast fringes on all three baselines without 
FINITO.  

We also downloaded AMBER data of Betelgeuse obtained 
on 2006 February 10 (Program ID: 60.A-9054A) from the ESO data archive.  
These data were taken with two ATs in the E0-G0-16~m configuration using the 
$K$-band medium-resolution mode (MR\_K) without FINITO.  
The wavelength range between 2.1 and 2.2~\micron\ was covered 
with a spectral resolution of 1500.  
Since there are no strong spectral features in this wavelength region, 
these \MRK\ data approximately sample the continuum.  
In both runs in 2008 and 2006, Sirius ($\alpha$~CMa, A1V, $K$ = $-1.4$) 
was observed 
for the calibration of the interferometric data of Betelgeuse.  
We adopted an angular diameter of 5.6~mas for Sirius given by 
Richichi \& Percheron (\cite{richichi05}).  
Observations of stars with known angular diameters and no asymmetry 
are necessary to evaluate the so-called interferometer transfer 
function, which represents the instrumental and atmospheric effects 
on visibility and phase measurements, and to calibrate 
interferometric data of a science target.  
A summary of the observations is given in Table~\ref{obslog}.  

For the reduction of the AMBER data, we used amdlib ver.2.2\footnote{
Available at http://www.jmmc.fr/data\_processing\_amber.htm}, 
which is based on the P2VM algorithm (Tatulli et al. \cite{tatulli07}).  
We split each data set of Betelgeuse and Sirius into five or six subsets 
with each subset containing 500 frames and 
derived the visibilities, DPs, and CPs, as well as spectra, from 
each subset.  
One of the parameters in the reduction with amdlib is the frame selection 
criterion.  
For each subset, 
we checked for a systematic difference in the results by taking 
the best 20\%, 40\%, 60\%, 80\%, and 100\% of all frames in terms of the fringe 
S/N ratio.  
We found out that the visibilities obtained from the first two subsets of 
the Betelgeuse data \#2 and the Sirius data taken in 2008 show a significant 
dependence on the selection criteria (the more frames are included, the lower 
the visibility becomes), while the visibilities from the other subsets 
do not show such a dependence.  Furthermore, this dependence occurs 
only on the 16~m and 48~m (E0-G0 and E0-H0) baselines, while the 
visibilities on the 32~m (G0-H0) baseline are not affected by the 
selection criterion in any subset.  The vibration of the AT at the E0 
station is very likely to be the cause of this problem, because it appears 
only on the baselines using the E0 station, and vibration always lowers 
visibility.  Therefore, we dropped these subsets affected by the 
AT vibration for the derivation 
of the visibilities, DPs, and CPs on the 16~m and 48~m baselines, 
while we used all subsets for the observables on the 32~m baseline.  
For the subsets not affected by the vibration, 
the reduction with different selection criteria 
does not lead to a significant systematic difference in the results.  
The selection of fewer frames only results in larger errors in the final 
results, while the inclusion of frames with very poor S/N ratios 
produces spurious results at a few wavelengths. 
Therefore, we included 80\% of all frames in the subsets not affected 
by the vibration.  
For the AMBER \MRK\ data, we did not find a signature of vibration, so 
we used all subsets with 80\% of frames included.  
Currently, AMBER data taken in the \HRK\ mode are affected by the Fabry-Perot 
effect caused by the InfraRed Image Sensor (IRIS), which stabilizes the image 
motion (see Fig.~6 in Weigelt et al. \cite{weigelt08}).  
The IRIS Fabry-Perot effect is seen as time-dependent high-frequency
beating in the raw visibilities, DPs, and CPs of Betelgeuse and Sirius plotted 
as a function of wavelength.  However, fortunately, it is mostly removed by 
dividing the data of Betelgeuse with that of Sirius, and the IRIS
beating is barely discernible in the calibrated visibilities, DPs, and CPs of 
Betelgeuse.  This is because the data of Betelgeuse and Sirius were taken 
close in time.  In other cases, the IRIS beating degrades the final data 
significantly.  

While the visibilities and DPs on the shortest baseline reduced from the 
data taken with a spectral resolution of 12000 are of sufficient 
quality, the visibilities and DPs on the longer baselines and CPs turned 
out to be noisy.  Therefore, we improved the S/N ratio for these observables 
by binning the 
data (object, dark, sky, and P2VM calibration data) in the spectral 
direction.  
For the visibilities and DPs on the middle baseline, binning 
with a box car filter with a width of three pixels turned out to be 
sufficient.  
This results in a spectral resolution of 8000 instead of 12000 achieved 
with the two-pixel sampling.  For the observables on the longest baseline 
and CPs, it was necessary to bin the data 
with five pixels, corresponding to a resolution of 4800, to obtain 
reasonable S/N ratios.  
As shown below, the individual CO lines can still be resolved with 
these lowered spectral resolutions.  

The errors of the resulting visibilities, DPs, and CPs were estimated 
from the standard deviation among the results obtained from five or 
six subsets.  The errors 
of the calibrated observables were computed from such errors in the 
data of Betelgeuse and Sirius.  
Since we have only one data set of Sirius, it is impossible to estimate 
the systematic error in the transfer function.  
Therefore, to account for this error source, 
we added a systematic error of 5\% to the above errors.

In both runs in 2008 and 2006, 
Sirius served not only as an interferometric calibrator but also as a 
spectroscopic standard star.  We attempted to remove telluric lines as 
much as possible by dividing the spectra of Betelgeuse with 
that of Sirius, although the difference in air mass for the 2008 data 
did not allow us to achieve this perfectly. 
The telluric lines identified in the spectrum of Sirius were used 
for wavelength calibration.  As a template of the telluric lines, 
we convolved the atmospheric transmission spectra from Wallace \& Hinkle 
(\cite{wallace96}) to match the resolutions of the 
HR\_K and MR\_K modes of AMBER. 
The uncertainty in wavelength calibration is 
$\sim \!\! 6 \times 10^{-5}$~\micron\ ($\sim$8~\KMS) for the \HRK\ 
observations and $\sim \! \! 5 \times 10^{-4}$~\micron\ ($\sim$70~\KMS) 
for the \MRK\ observations.  
The wavelength scale was then converted to the 
heliocentric frame using the IRAF\footnote{IRAF is distributed by the 
National Optical Astronomy Observatories, which are operated by the 
Association of Universities for Research in Astronomy, Inc., under 
cooperative agreement with the National Science Foundation.} task 
RVCORRECT.

\subsection{VINCI}

To discuss temporal variations in the $K$-band visibility, we also 
downloaded interferometric data of Betelgeuse 
taken with VINCI (VLT INterferometer Commissioning Instrument) 
from the ESO archive (Program ID: 60.A-9222A).  
These data were obtained as part of the commissioning of the instrument. 
A detailed description of the instrument is given in 
Kervella et al. (\cite{kervella00}).  
As summarized in Table~\ref{obslog_vinci}, 
the VINCI observations of Betelgeuse occurred on eight nights between 
2001 and 2003 using two 40~cm siderostats in the E0-G0-16m and B3-C3-8m 
configurations (both configurations 
lie at $+$71\degr\ from North ($+$90\degr\ = East) on the ground).  
A number of calibrators were observed 
on these nights (6--35 calibrator measurements on each night), 
as listed in Table~\ref{cal_vinci}.  

We used the VINCI data reduction software ver.3.0\footnote{Available 
at http://www.jmmc.fr/data\_processing\_vinci.htm} (Kervella et al. 
\cite{kervella04}) to derive visibility.  The interferometer transfer 
function was computed from all calibrator measurements taken during 
a given night, and the mean of these transfer function values was 
used to obtain the calibrated visibilities of the science target.  
The error of each calibrated  visibility was derived from the 
statistical error in each measurement of the science 
target given by the reduction software and the error in the transfer 
function.  This latter error results from the statistical error 
in each calibrator measurement and the standard deviation of the 
transfer function values obtained on the given night.  
The VINCI data reduction software computes visibility using 
two different algorithms: Fourier transform and wavelet transform. 
For the data presented here, the calibrated visibilities 
derived with two methods agree well, and we only give the results 
obtained with the wavelet transform in Table~\ref{obslog_vinci}. 
The errors in the calibrated visibilities are typically 1--3\%. 
Unlike AMBER, VINCI observations were made with the $K$-broadband 
filter covering from 2 to 2.4~\micron.  Therefore, we computed 
the effective wavelength using the VINCI transmission presented 
in Wittkowski et al. (\cite{wittkowski04}) and the spectrum of 
Betelgeuse observed with the Stratoscope II (detector B) 
by Woolf et al. (\cite{woolf64}).  The resulting effective 
wavelength, 2.175~\micron, was used for calculating the 
spatial frequency for each observation.

\section{Results}
\label{sect_results}

Figure~\ref{obsres} shows the calibrated visibilities, DPs, and CPs observed 
toward Betelgeuse as a function of wavelength.  
The visibilities and DPs on the middle and longest baselines, as well 
as the CPs were derived from the binned data (spectral resolutions of 
8000 and 4800), while the results on 
the shortest baseline were derived from the data with 
a spectral resolution of 12000.  
The figure reveals the detection of interferometric signals even on the 
longest baseline (48~m).  
This marks the highest spatial resolution (9~mas) obtained for Betelgeuse, 
corresponding to nearly five resolution elements over its stellar disk.  
The visibilities, DPs, and CPs derived from the two data sets show 
mostly the same spectral features, 
demonstrating that the observed features are real despite the low fringe 
contrast.  
There seems to be a difference in visibility level for the 
E0-G0-16m baseline.  
However, given the uncertainties in the calibrated visibilities shown in 
the figure, this discrepancy is marginal.  
We note that the CPs near the CO band head 
between 2.294 and 2.296~\micron\ are much noisier---errors of 
50--100\degr ---than at the other wavelengths even in the binned data.

\subsection{Continuum: $\lambda < 2.293$~\micron}
\label{subsect_res_cont}

The observed spectrum of Betelgeuse below 2.293~\micron\ shows only 
several weak atomic and molecular absorption features, as identified 
in Fig.~\ref{obsres}a.  There are also subtle signatures in the observed 
visibilities corresponding to these features.  In particular, the 
effects of the Ti (+ HF) feature at 2.29~\micron\ can be seen in the 
visibilities on all three baselines and also 
possibly in the CPs.  These weak lines form in the deep photospheric 
layers and can be used for testing photospheric models, which we 
plan in a future paper.  

The spectral resolution of the HR mode of AMBER allows us to select 
continuum points which are not affected by these lines.  
Since the visibilities on the longest baseline were 
derived with the five-pixel binning, 
we also derived the visibilities on the shorter baselines 
and the spectra from the data binned with five pixels.  
Then we selected 37 continuum points in the spectra, avoiding the lines.  
In Fig.~\ref{vis_continuum}, we plot the visibilities observed at these 
continuum points below 2.293~\micron\ as a function of spatial frequency.  
Also plotted are the AMBER \MRK\ data between 2.1 and 2.2~\micron, 
VINCI data, 
$K$-band data obtained at the Infrared Optical Telescope Array 
(IOTA) by Perrin et al. (\cite{perrin04}), and 
the $K$-band measurements by Dyck et al. (\cite{dyck92}).  
The visibilities derived from the AMBER MR\_K data agree very well with 
the result obtained by Hern\'andez \& Chelli (\cite{hernandez07}) using 
not only amdlib but also another algorithm based on the Fourier transform.  
The errors in the visibilities from the \MRK\ data are $\sim$6\%.  
The DP between 2.1 and 2.2~\micron\ derived from these data is zero 
within a measurement error of $\sim$5\degr, which is no surprise, given 
the absence of strong spectral features in this wavelength range.  
While the AMBER HR\_K and MR\_K data measure the continuum almost free 
from the effects 
of molecular/atomic features, the VINCI and IOTA data as well as the 
measurements of Dyck et al. (\cite{dyck92}) 
were taken with a broad band filter spanning the entire $K$ band, 
which includes the molecular absorption features due to CO, \HOH, and CN 
and atomic lines.  
However, since the CO and \HOH\ bands appear only at limited wavelengths 
(CO: longward of 2.3~\micron, \HOH : either edge of the $K$ band), 
and CN and atomic features are weak in Betelgeuse, 
the total $K$-band flux is dominated by the continuum.  
Therefore, the data taken with the $K$ broad-band filter represent the 
visibilities in the continuum in first approximation, and this is why 
they are included in the discussion below.

To derive the angular size of the object in the continuum, 
we fitted the observed visibilities with a uniform disk and a 
limb-darkened disk.   
Uniform-disk fitting to the AMBER continuum data in 2008 and 2006 
results in $43.19 \pm 0.03$~mas (reduced $\chi^2$ = 4.6) and 
$42.69 \pm 0.01$~mas (reduced $\chi^2$ = 0.01), respectively, 
while the fitting to all the data (AMBER, IOTA, and VINCI) 
results in $43.16 \pm 0.03$~mas (reduced $\chi^2$ = 3.3).  
Fitting to the AMBER \HRK\ data 
with a power-law-type limb-darkened disk (Hestroffer et al. 
\cite{hestroffer97}) results in a limb-darkened disk diameter of 
$43.56 \pm 0.06$~mas and a limb-darkening parameter of 
$(1.2 \pm 0.07) \times 10^{-1}$ (reduced $\chi^2$ = 3.44), 
while fitting to all the data results in a limb-darkened disk 
diameter of $43.50 \pm 0.04$~mas and a limb-darkening parameter of 
$(9.0 \pm 0.4) \times 10^{-2}$ (reduced $\chi^2$ = 2.5).  
These results agree well with the limb-darkened disk diameter 
and the limb-darkening parameter derived by Perrin et al. (\cite{perrin04}) 
from the $K$-broadband IOTA data.  
Since a uniform disk has a limb-darkening parameter of 0, the fit suggests 
a small limb-darkening effect in the $K$-band continuum.  
Limb-darkened disk fitting to the AMBER \MRK\ data did not give 
a meaningful result, because the data are located in one narrow spatial 
frequency range.  
The fit to the AMBER \HRK\ data in 2008 
shows that the 16, 32, and 48~m baselines correspond 
to the second, third, and fifth lobes.  
Although the longest baseline falls near the fourth null (i.e., 
between the fourth and fifth lobes), 
the detection of fringes in the fifth lobe is corroborated by the 180\degr\ 
CPs measured in the continuum below 2.293~\micron\ 
(Fig.~\ref{obsres}e). 
This is because CP is the sum of the phases on the three 
baselines (E0-G0, G0-H0, and H0-E0), that is, CP = 180\degr\ (2nd lobe) + 
0\degr\ (3rd lobe) + 0\degr\ (5th lobe) = 180\degr.  
While the above fit shows the AMBER \HRK\ data cannot be perfectly 
fitted with a uniform disk or limb-darkened disk (i.e., $\chi^2 > 1$), 
Fig.~\ref{vis_continuum} suggests that the deviation from the fitted curves 
is not drastic at the time of the observations.  
This means that the spatial scale of inhomogeneities in the $K$-band 
continuum is even smaller than 
the resolution of the longest baseline (9~mas) and/or the 
contrast of the inhomogeneities is small.  
This picture is also supported by the result that the CPs measured in the 
continuum are 180\degr\ within the errors, 
as expected from the above fit with a uniform disk or a limb-darkened disk.

The data in the second and third lobes were taken at seven different 
epochs as given in Fig.~\ref{vis_continuum}.  
Therefore, these data points contain information about the 
effects of time-dependent surface inhomogeneities 
on the $K$-band visibility in the continuum.  
Recent 3-D hydrodynamical convection simulations for RSGs by 
Chiavassa et al. (\cite{chiavassa08a}) and Ludwig \& Beckers (\cite{ludwig08}) 
predict significant temporal variations in visibility beyond the first null 
due to ever-changing inhomogeneities.  
For example, Chiavassa et al. (\cite{chiavassa08a}) present the 
amplitude of such variations in the 2.2~\micron\ visibility in the second 
and third lobes predicted for Betelgeuse, which are plotted by 
the dotted lines in Fig.~\ref{vis_continuum} 
(the model visibility at 2.1~\micron\ presented in 
Ludwig \& Beckers (\cite{ludwig08}) also shows a similar behavior).  
However, most of the observational data points in the second and third lobes 
are located near the upper boundary or at least in the upper half of the 
predicted range, instead of equally distributed above and below the 
center of the range.  This means that the model predicts the visibility 
to be systematically lower than the observations.  
The observed data in the second lobe, 
where we have AMBER, VINCI, and IOTA data from five epochs, 
show only a modest scatter around the limb-darkened-disk fit.  
The data on the third lobe do not show large 
variations, either, although we have data from only two epochs.  
Therefore, the AMBER, VINCI, and IOTA observations 
imply that the 3-D convection simulation of Chiavassa et al. 
(\cite{chiavassa08a}) predicts too large deviations 
(presumably due to too strong inhomogeneities) from 
the limb-darkening disk which fits the observed data reasonably 
in the $K$-band continuum.

However, given that only our AMBER \HRK\ data probes the very 
high-spatial frequency regime and the number of the observation epochs is 
not yet very large, we cannot conclude whether or not the 
visibility at these high spatial frequencies shows little deviation from a 
uniform-dirk or limb-darkened disk all the time.  
Also, only our AMBER observations have 
measured CP, whose deviations from zero or $\pi$ would represent a clear 
signature of asymmetry.  
The visibilities of Betelgeuse observed with the $H$-broadband filter 
with IOTA show more deviation from a uniform disk than in the $K$ band 
continuum, and non-zero CPs are also detected in the third and fourth lobes 
(Haubois et al. \cite{haubois06}).  
While these results reveal inhomogeneous structures seen 
with the $H$-broadband filter (including the absorption features due to 
molecules such as \HOH, CO, and OH), it is not 
yet clear whether the amplitude of the deviation is consistent with 
the current convection simulation models, 
because the $H$-band IOTA data represent only one epoch.  
Further spectro-interferometric monitoring observations in the $K$ and 
$H$ bands at such high 
spatial frequencies as obtained here will enable us to study the 
spatial and time scale of inhomogeneities in the continuum and provide 
stronger observational tests for the 3-D simulations of convection in 
RSGs.  Particularly, multi-epoch observations at the same spatial 
frequencies will provide a direct test for the visibility fluctuation 
predicted by the simulations.

Recently, Townes et al. (\cite{townes09}) have reported a monotonic 
decrease by 15\% in the 11~\micron\ diameter of Betelgeuse in the past 15 
years, from 1993 to 2009.  Its origin is unclear.  
We examined a possible long-term variation 
in the $K$-band uniform-disk diameter in the interferometric 
observations in the literature and the VINCI data.  
The uniform-disk diameters derived from the first-lobe data in the past 
are $44.2\pm 0.2$~mas (Aug.-Sep. 1990, Dyck et al. \cite{dyck92}) 
and $43.26 \pm 0.04$~mas (Nov. 1996, Perrin et al. \cite{perrin04}).  
We fitted the VINCI data in the first lobe taken in Dec. 2002 and 
Jan. 2003 with a uniform disk model computed with the $K$-band 
transmission of VINCI and the spectrum of Betelgeuse, which were 
used for computing the effective wavelength.  
The resulting uniform-disk diameter, $42.02 \pm 0.4$~mas, 
agrees very well with the $42.06 \pm 0.4$ 
obtained by Meisner (priv. comm.) from the same VINCI data but using a 
different algorithm based on coherent integration (Meisner \cite{meisner03}).  
The error includes the uncertainty in the determination of the effective 
wavelength in VINCI observations (Meisner, priv. comm.), which we also 
added to the error in our uniform-disk diameter.  
Therefore, the $K$-broadband uniform-disk diameter shows a decrease by 
$5\pm 2$\% from 1990 to 2003, much less pronounced than at 11~\micron. 
The continuum uniform-disk diameters of 42.69 and 43.19~mas obtained from the 
AMBER data taken in 2006 and 2008 might indicate an increase in the 
angular size after 2003.  
However, these diameters obtained from the data beyond the first null 
can be affected by inhomogeneities (although not very strong), 
which makes the apparent increase in the angular size inconclusive.

\subsection{CO first overtone lines: $\lambda > 2.293$~\micron}

\begin{figure*}[!th]
\resizebox{\hsize}{!}{\rotatebox{0}{\includegraphics{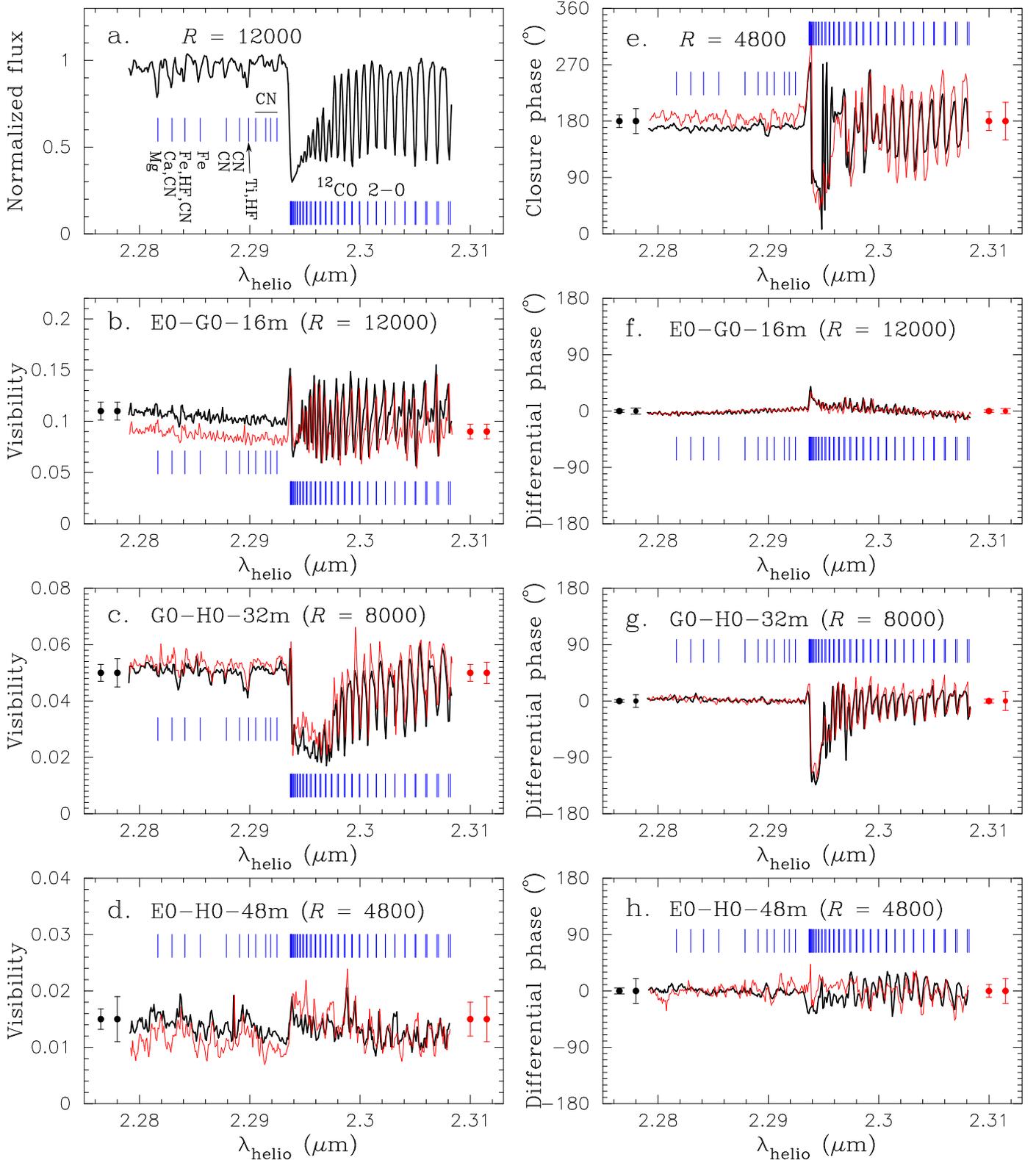}}}
\caption{AMBER observations of Betelgeuse.  
In all panels except for {\bf a}, the black and red lines 
represent the data set \#1 and \#2, respectively.  
The spectrum shown in the panel {\bf a} was derived from the 
merged data.  
The positions of the CO 
first overtone lines as well as other atomic and molecular lines 
are marked with the ticks.  
In the panels except for {\bf a}, 
two error bars near the left ordinate represent 
the typical errors in the continuum (left) and in the CO lines (right) for 
the data set \#1.  
The errors for the data set \#2 are shown near the right ordinate 
in the same manner.  
The error in the normalized spectrum is 0.5\% and 1\% in the continuum 
and in the CO lines, respectively.  
The wavelength scale is in the heliocentric frame.  
{\bf a:} Normalized flux.  
{\bf b--d:} Visibilities observed on the E0-G0-16m, G0-H0-32m, and 
E0-H0-48m baselines.  The visibilities on the middle and longest baselines are 
binned with a box car filter with widths of three and five pixels, 
respectively.  
{\bf e:} Closure phases with five-pixel binning.  
Note that the errors in CP near the CO band head between 2.294 and 
2.296~\micron\ are as large as 50--100\degr.
{\bf f--h:} Differential phases observed on the E0-G0-16m, G0-H0-32m, and 
E0-H0-48m baselines.  The DPs on the middle and longest baselines are 
binned with three and five pixels, respectively.  
}
\label{obsres}
\end{figure*}

\begin{figure*}
\sidecaption
\resizebox{\hsize}{!}{\rotatebox{-90}{\includegraphics{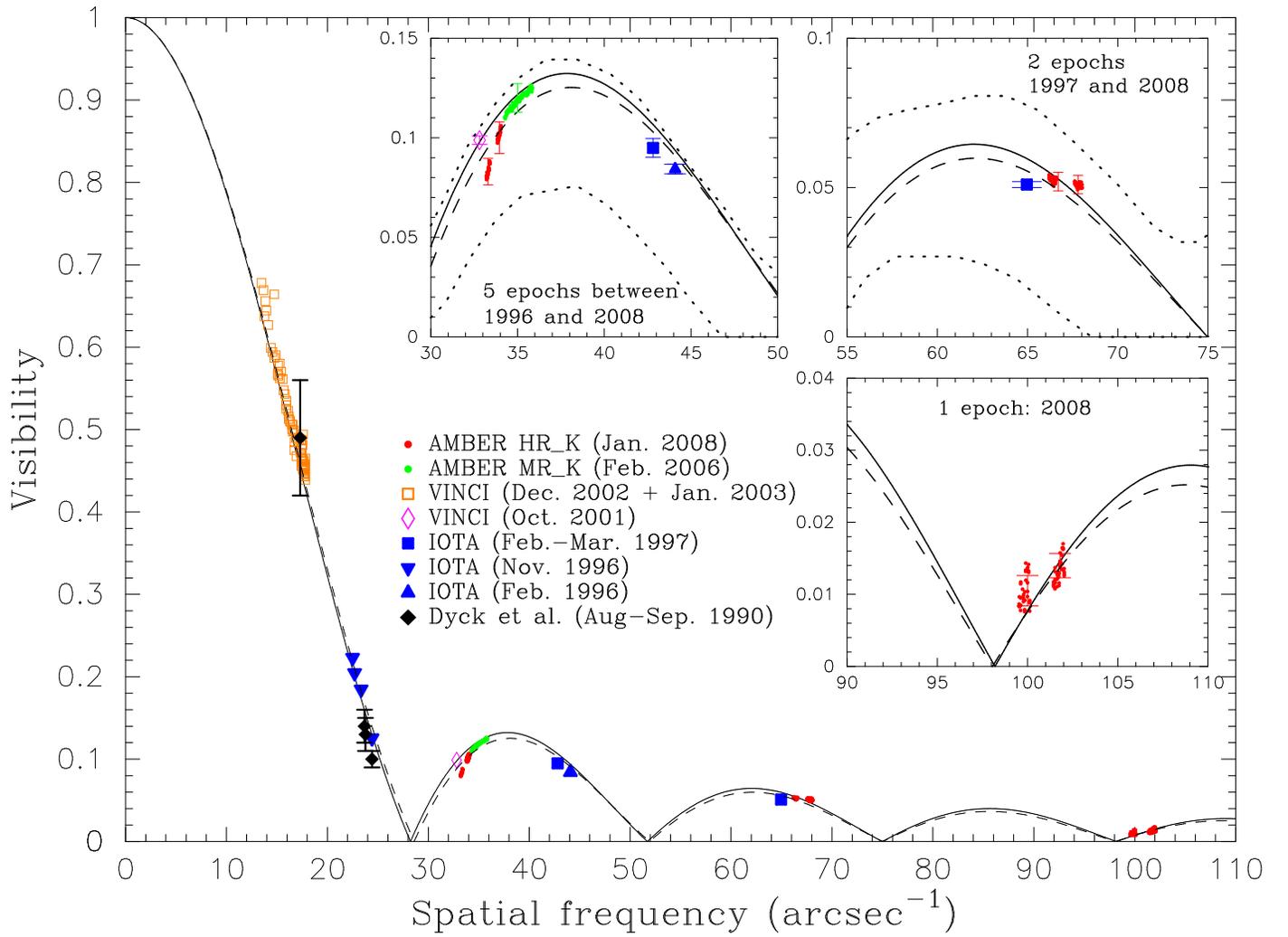}}}
\caption{$K$-band continuum/broadband visibilities of Betelgeuse 
plotted as a function 
of spatial frequency.  The insets show enlarged views of the second, third, 
and fifth lobes.  
The error bars of the single AMBER data points are exemplarily shown in the 
insets.  The errors of the VINCI and IOTA data are also shown in the insets.  
The solid and dashed lines represent the visibilities for a uniform disk with 
a diameter of 43.19~mas and for a limb-darkened disk with a diameter of 
43.56~mas 
and a limb-darkening parameter of 0.12 (power-law-type limb-darkened disk 
of Hestroffer (\cite{hestroffer97}), respectively.  
The dotted lines represent the full amplitude of the 
variations in the 2.22~\micron\ visibility due to time-dependent 
inhomogeneous structures predicted by the 3-D convection simulation of 
Chiavassa et al. (\cite{chiavassa08a}). 
}
\label{vis_continuum}
\end{figure*}

\begin{figure}
\resizebox{\hsize}{!}{\rotatebox{0}{\includegraphics{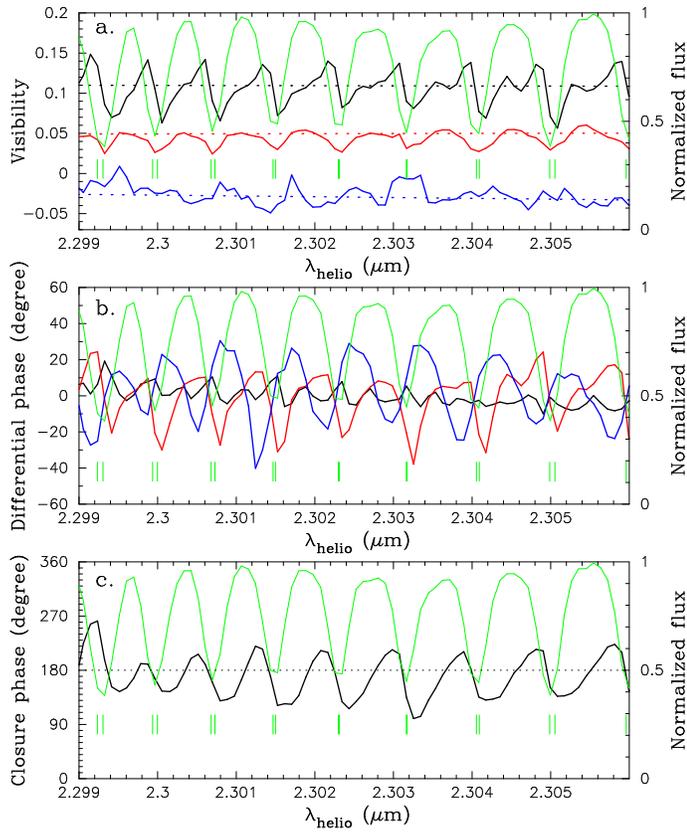}}}
\caption{Visibilities ({\bf a}), differential phases ({\bf b}), 
and closure phases ({\bf c}) observed in the CO first overtone lines 
toward Betelgeuse.  
In the panels {\bf a} and {\bf b}, the black, red, 
and blue lines represent the visibilities or differential phases observed 
on the E0-G0-16m, G0-H0-32m, and E0-H0-48m baselines, respectively. 
In the panel {\bf a}, the visibility obtained on the longest baseline 
(blue) is scaled by a factor of six and shifted downward by 0.1 for the 
sake of visual clarity.  
The black, red, and blue dotted lines represent the continuum visibilities 
of a uniform disk with 43.19~mas for the E0-G0-16m, G0-H0-32m, and 
E0-H0-48m baselines, respectively.  
In the panel {\bf c}, the observed CP is shown by the black solid line.  
The dotted line represents CP = $\pi$, which is observed in the continuum.  
The visibilities and DPs on the middle and longest baselines were 
derived from the data binned with three and five pixels, respectively.  
The CP was derived from the data binned with five pixels.  
In each panel, the normalized flux (without binning) 
is overplotted in green to show the behavior of the visibilities and phases 
within individual CO lines.  
The positions of the CO lines are marked with the vertical ticks. 
The wavelength scale is in the heliocentric frame.  
A color version of this figure is available in the electronic edition. 
}
\label{obsresCO}
\end{figure}

Figure~\ref{obsres} reveals salient signatures of the CO first overtone 
lines in the visibilities, DPs, and CPs longward of 2.293~\micron.  
Figure~\ref{obsresCO} shows the enlarged views between 
2.299 and 2.306~\micron\ (only the data set \#1 is plotted for visual 
clarity), where the observed spectrum and the line positions 
are overplotted in each panel to show 
the shape of the visibilities, DPs, and CPs within the CO lines.  
Each absorption feature consists of two transitions with 
high and low rotational quantum numbers $J$.  But for simplicity, 
we refer to such an absorption feature with high and low $J$ as a ``line'' 
in the present work. 
Comparison of the observed line positions with the laboratory data results 
in a heliocentric velocity of 24~\KMS.  
This value agrees with the velocity range 17--27~\KMS\ derived from 
the visible and IR atomic and molecular lines as well as the mm 
CO lines (e.g., Brooke et al. \cite{brooke74}; 
Huggins \cite{huggins87}; Smith et al. \cite{smith89}; 
Huggins et al. \cite{huggins94}; Ryde et al. \cite{ryde99}).  

Figure~\ref{obsresCO}a shows the visibilities observed on the three 
baselines.  Particularly surprising is that the visibility observed 
within a given CO line on the shortest baseline ($\sim$16~m) is 
anti-symmetric with respect to the line core (i.e., ``$\sim$''-shaped).  
While the visibility on the 
middle baseline ($\sim$32~m) is roughly symmetric, 
the visibility on the longest baseline ($\sim$48~m) is asymmetric 
with the peak slightly redshifted with respect to the line core in most 
cases.  
These results mean that Betelgeuse appears different in the blue and red 
wings of the CO lines and that 
the blue and red wings originate 
in spatially distinct regions differing in size and/or shape. 
The observed DPs and CPs show remarkable non-zero and non-$\pi$ values, 
as large as DP = $-130$\degr\ (Fig.~\ref{obsres}g) or CP = 180\degr $\pm$ 
90\degr\ (Figs.~\ref{obsres}e and \ref{obsresCO}c).  
These non-zero and non-$\pi$ DPs and CPs confirm that the blue and red wings 
of the CO lines originate in spatially distinct regions.  
On the other hand, the observed DPs and CPs are nearly zero near the 
line core, which means that the star appears symmetric.

One might suspect that such blue-red asymmetry in visibilities, DPs, and CPs 
within 
the CO line profiles can be explained by the fact that one CO absorption 
feature, which appears to be a single line, is a blend of two transitions with 
low and high $J$.  If two lines with different excitation potentials form at 
different regions over the stellar surface, it may make the star appear 
different in the blue and red wings and cause the blue-red asymmetry.  
However, the relative positions of the high 
and low $J$ lines swap at 2.3032~\micron : shortward of this wavelength, 
the high $J$ lines appear blueward of the low $J$ lines, while above 
2.3032~\micron, the high $J$ lines appear redward of the low $J$ lines.  
Still, the same asymmetry is observed in all CO lines, whether shortward 
or longward of 2.3032~\micron.  Therefore, the observed asymmetry in the 
visibility, DPs, and CP cannot be explained by the blend of two transitions 
with high and low $J$. 

Stellar rotation or spherically expanding/infalling flows are also unlikely 
to be the cause of the blue-red asymmetry in visibilities and phases.  
The projected photospheric rotational velocity of 
Betelgeuse is small, $v \sin i \approx 2$--5~\KMS\ 
(Uitenbroeck \cite{uitenbroek98}; Harper \& Brown \cite{harper06}).  
Moreover, the rotational axis inferred from the chromospheric emission lines 
lies at $\sim$65\degr\ (Harper \& Brown \cite{harper06}), which is very close 
to the position angle of the projected baselines of our AMBER observations.  
In this case, rotation can have no noticeable effects on visibilities and 
phases.  
In spherically expanding/infalling flows, 
the velocity in the line of sight changes as a 
function of the angular distance from the stellar disk center.  
Therefore, the photons at different wavelengths (= different velocities) within a 
line profile originate in annular regions with different sizes, 
which makes the star appear different in the blue and red wings. 
However, obviously, such spherically symmetric models 
cannot explain the observed non-zero/non-$\pi$ DPs and CPs.  

The observed blue-red asymmetry in visibility, DP, and CP 
may be explained by an inhomogeneous velocity field in the atmosphere, 
in which upwelling and downdrafting CO gas exists in spatially 
distinct regions.  In spectroscopic 
analyses using 1-D model atmospheres, such a non-thermal velocity field is 
empirically incorporated as macroturbulence, which manifests itself as 
a broadening of spectral lines 
(``macro'' means the spatial scale of the non-thermal gas motion is 
larger than the length of the unit optical depth of photons). 
In the next section, we examine whether the observed visibilities, DPs, and 
CPs can be explained by such an inhomogeneous velocity field.

\section{Modeling}
\label{sect_modeling}

To characterize the inhomogeneous velocity field and the properties of 
the CO gas in the atmosphere of Betelgeuse, we constructed the following 
patchy two-layer model, in which the star is surrounded by the inner 
and outer CO layers.  
The star was assumed to be a blackbody of an effective temperature 
of 3600~K based on the value derived by Perrin et al. (\cite{perrin04}). 
The inner layer represents the 
photosphere---the region included in 1-D photospheric 
models, usually spanning from continuum optical depths of 
$\sim \! 10^{1}$--$10^{2}$ to $\sim \! 10^{-5}$--$10^{-6}$.  
The outer layer represents the MOLsphere.  
Given the presence of dense \HOH\ gas in the MOLsphere suggested for 
Betelgeuse 
from spectroscopic and interferometric observations 
(Perrin et al. \cite{perrin04}, \cite{perrin07}; Ohnaka \cite{ohnaka04a}; 
Tsuji \cite{tsuji06}), it is plausible 
that there is also a significant amount of CO in the MOLsphere.  

\begin{figure}
\resizebox{\hsize}{!}{\rotatebox{0}{\includegraphics{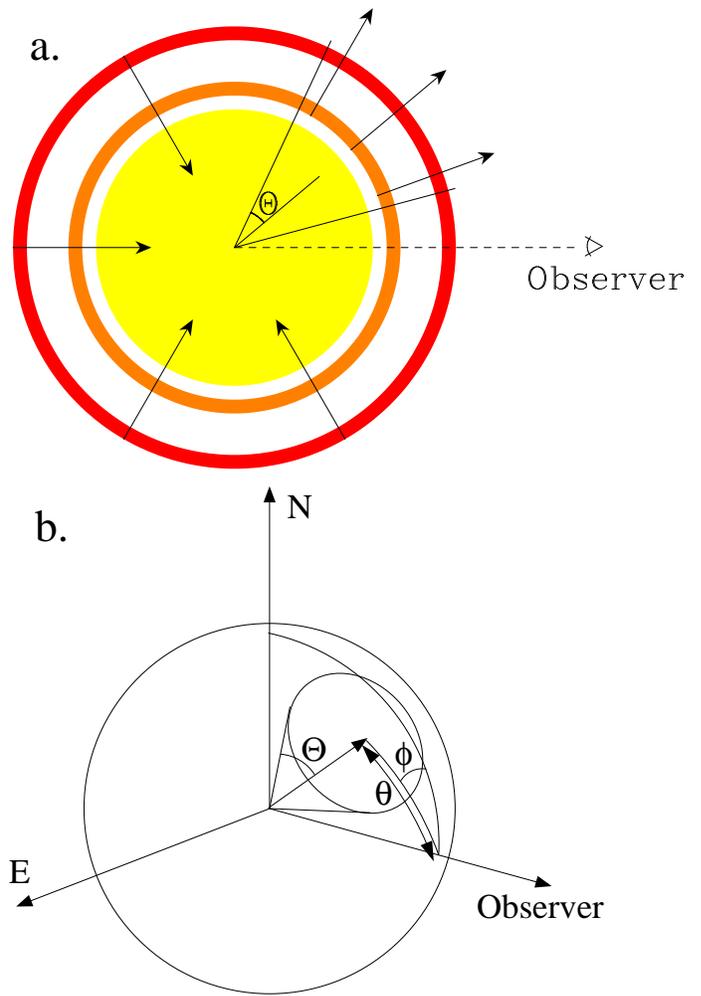}}}
\caption{
Schematic view of our patchy model for Betelgeuse. 
{\bf a:} Cross section of the model.  The star is surrounded by two layers 
with an inhomogeneous velocity field.  
{\bf b:} 3-D view of the model.  The N- and E-axes define the plane of 
the sky.  Only one layer is drawn for the sake of visual clarity.  
Within the cone specified by a half-opening angle of $\Theta$ and 
the position ($\theta$, $\phi$) with respect to the observer, the CO gas is 
assumed to be moving radially outward (or inward), while it is moving radially 
inward (or outward) in the remaining region.  
}
\label{alfori_model_schematic}
\end{figure}

The basic picture of our model is depicted in 
Fig.~\ref{alfori_model_schematic}.  
This two-layer model is similar to that described in 
Ohnaka (\cite{ohnaka04b}), but we introduced two modifications. 
Firstly, the geometrical thickness of each 
layer is assumed to be very small compared to its 
radius, and therefore, the two layers are detached from each other.  
This simplifies the computation of the line opacity in the presence 
of a velocity field as described below.   
Secondly, while the column density and temperature of each layer were assumed 
to be constant over the stellar surface, 
we incorporated the following inhomogeneous velocity field: 
CO gas is moving radially outward (or inward) with a velocity of \VFLOW\ 
in one patch, while it is moving in the opposite direction 
at the same velocity in the remaining region.  We assumed the same 
velocity field for two layers. 
As Fig.~\ref{alfori_model_schematic}b illustrates, 
we define such a patch as a cone.  
Its half-opening angle ($\Theta$) characterizes the size of the patch, 
while its position is specified by two angles, $\theta$ and $\phi$, 
which define the vector connecting the center of the patch 
and the center of the star.  

We do not know a priori the 
actual number and shape of patches (if any) on the stellar surface, and 
the present data are insufficient to derive the inhomogeneous surface 
pattern uniquely.  
We assume only one patch in our modeling to keep the number of free 
parameters as small as possible.  
The aim of our modeling is not to derive the actual inhomogeneous 
surface structure but to see if there is indeed a model with 
an inhomogeneous velocity field which can explain the observed blue-red 
asymmetry in visibility, DP, and CP within the CO lines.  

To decrease the number of free parameters, we fixed the CO column density 
and radius of the inner layer as follows.  
A spherical photospheric model with the molecular opacities due to 
CO, TiO, \HOH, OH, and SiO incorporated based on the opacity sampling 
(Ohnaka, in prep.) was computed using the stellar parameters of Betelgeuse 
given in Tsuji (\cite{tsuji06}).  
This photospheric model gives a CO column density of 
$5 \times 10^{22}$~\PERSQCM, which we adopted for the inner layer.  
We assumed the radius of the inner CO layer to be a half of 
the geometrical thickness of this photospheric model ($\sim$0.1~\RSTAR), 
that is, 
a radius of 1.05~\RSTAR.  While this choice is ambiguous, it turned out 
not to affect the result significantly, as far as the radius 
is smaller than $\sim$1.1~\RSTAR.  
It is necessary to include the microturbulence, which represents the 
non-thermal gas motion on a spatial scale smaller than the length of the 
unit optical depth of photons.  
The microturbulent velocity in the photosphere of Betelgeuse derived 
from spectroscopic analyses ranges from 4 to 6~\KMS\ (Lambert et al. 
\cite{lambert84}; Tsuji et al. \cite{tsuji94}; Tsuji \cite{tsuji06}).  
We adopted a microturbulent velocity of 5~\KMS\ for both the inner and 
outer CO layers in our modeling.  

For the outer layer, we fixed its radius and temperature based on the 
previous spectroscopic and interferometric studies of the MOLsphere.  
The radius of the MOLsphere of Betelgeuse measured from the near- and 
mid-IR \HOH\ features is 1.45~\RSTAR\ (Ohnaka \cite{ohnaka04a}), 
1.3~\RSTAR\ (Tsuji \cite{tsuji06}), and 1.31--1.43~\RSTAR\ 
(Perrin et al. \cite{perrin07}).  We tentatively adopted 1.45~\RSTAR\ for 
our modeling.  
The temperature of the \HOH\ MOLsphere derived by the above authors range 
from 1500 to 2250~K.  A gas temperature of 1800~K, which lies roughly 
in the middle of this range, was adopted for the outer CO layer.  
Therefore, the free parameters of our model are the temperature of the 
inner CO layer (\TIN), the CO column density of the outer CO layer 
(\NCOOUT), the velocity of the CO gas motion (\VFLOW), 
and the position and size of the patch ($\theta$, $\phi$, and $\Theta$).

The intensity distribution for this patchy model was computed at each 
wavelength between 2.291 and 2.309~\micron\ by performing ray tracing 
along a number of lines of sight (see Ohnaka \cite{ohnaka04a}; 
\cite{ohnaka04b}).  
The only difference from these previous studies is the inclusion of 
the Doppler shift 
due to the velocity field in the calculation of the line opacity.  
Since each layer is geometrically thin, the velocity 
component along the line of sight within one layer can be approximated 
to be constant, which simplifies the computation.  
The CO line list of Goorvitch (\cite{goorvitch94}) was used for the 
calculation of the line opacity.  
The monochromatic 2-D complex visibility was calculated by taking the 
Fourier transform of the intensity and then spectrally convolved to 
match the resolution of our AMBER observations (12000, 8000, and 4800). 
The visibility amplitude and phase were 
derived from this spectrally convolved 2-D complex visibility, which 
was scaled with the uniform-disk diameter of 43.19~mas derived from 
the AMBER \HRK\ data in the continuum.  
CP was obtained as the sum of Fourier phases 
predicted for three baselines, 
while DP on a given baseline was computed by taking 
the difference in Fourier phase between in a CO line and in the continuum 
below 2.293~\micron.  
The correspondence between the sign of CP measured with AMBER and 
the positional offset in the plane of the sky is described 
in Kraus et al. (\cite{kraus09}).

Figure~\ref{alfori_model} shows a model with 
\TIN\ = 2250~K and \NCOOUT\ = $1 \times 10^{20}$~\PERSQCM, 
characterized by a large, upwelling spot ($\Theta$ = 60\degr, \VFLOW\ = 
10~\KMS) covering nearly a half of the apparent stellar disk ($\theta$ = 
40\degr, $\phi$ = 10~\degr, see also Figs.~\ref{alfori_images}b and 
\ref{alfori_images}d).   
Given the simple nature of the model, the overall agreement is reasonable, 
although there are still discrepancies between the model and the 
observed data.  
Particularly, the agreement is poor near the band head, which is 
discussed below.  Also, the visibility on the longest baseline as well as 
the CP predicted 
by the model is too smooth compared to the observed data (note that the 
spectral resolution of the model matches that of the binned data).  
Figure~\ref{alfori_images} shows the images predicted by 
this model at different wavelengths within a CO line ($R$(29) and $R$(72)). 
The figure illustrates that 
the velocity field makes the star appear different in the blue and red 
wings.  
This can cause the visibility to be anti-symmetric with respect to the line 
core as observed on the shortest baseline and also explains the 
observed asymmetric DPs and CPs.  
The stellar image at the line core is the sum of the blue- and 
red-shifted components.  This makes the star appear nearly centrosymmetric 
(Fig.~\ref{alfori_images}c) and results in DPs and CPs close to zero 
at the line core as observed.  

The agreement between the observed data and the model becomes poorer near 
the band head.  In particular, the deviations of DP from zero predicted 
for the middle and longest baselines are too small compared to the observed 
values.  
The reason why the DPs predicted by the model do not show significant 
deviation from zero is as follows.  
Since the lines are very crowded near the band head, the blue (or red) wing 
of one line overlaps with the red (or blue) wing of the adjacent lines. 
This means that 
the appearance of the star at a given wavelength in such a crowded region 
is the sum of the blue- and red-shifted components as in the case of 
the image at the core of an isolated line.  
As Fig.~\ref{alfori_images}c shows, the resulting image appears roughly 
centrosymmetric.  
This is the reason why the DPs predicted by the model do not show 
noticeable deviations from zero.  
A possible reason for the discrepancy near the band head 
is our assumption that the CO column density and temperature are 
uniform over the stellar surface.  
Inhomogeneities in CO column density and/or in temperature over the surface 
may reconcile the disagreement between the model and the observed data near 
the band head.  

The uncertainty ranges of \TIN, \NCOOUT, and \VFLOW\ derived from our 
modeling are 2000--2500~K, 
$5 \times 10^{19}$--$2 \times 10^{20}$~\PERSQCM, and 10--15~\KMS, 
respectively.  
We also computed models with \NCOOUT\ set to zero (i.e., no MOLsphere), but 
such models cannot reproduce the visibilities observed near the band head. 
The uncertainties in the position and size of the patch are very 
large: the ranges for $\Theta$, $\theta$, and $\phi$ are 
20\degr --80\degr, 20\degr --80\degr, and 0\degr --40\degr, respectively.  
However, these uncertainties in the position and size of the patch 
should not be taken at face value, because we assumed for simplicity that 
there is only one spot, which may not be the case.  
Also, we cannot conclude whether the CO gas in the patch is 
moving upward or downward.  We computed models in which the patch is moving 
downward and found out that there are parameter sets which are consistent 
with the observations.  
However, apart from these uncertainties in the actual 
surface pattern and the direction of the velocity field, 
our AMBER data and modeling suggest an inhomogeneous velocity field with 
amplitudes of $\sim$10--15~\KMS.

\section{Discussion}
\label{sect_discuss}

The amplitude of the velocity field suggested from our modeling, 10--15~\KMS, 
compares favorably with the macroturbulent velocities derived in the previous 
spectroscopic analyses.  
Macroturbulence manifests itself as the broadening of spectral lines 
in observed spectra.  
In spectral analyses using 1-D model atmospheres, 
macroturbulence is usually incorporated as an additional line broadening 
factor with a Gaussian distribution (i.e. $\propto 
\exp(-(\lambda/\lambda_{\rm macro})^2$)).  
The macroturbulent velocity---either 
as the Gaussian dispersion $V_{\rm macro}$ or the FWHM = 
$ 1.665 V_{\rm macro}$ corresponding 
to $\lambda_{\rm macro}$---is derived so that synthetic spectra reproduce 
observed line profiles.  
For Betelgeuse, macroturbulent velocities as high as 
20~\KMS\ (FWHM) were derived from the optical lines by Lobel \& Dupree 
(\cite{lobel00}) and Gray (\cite{gray00}), while smaller values of 
10--12~\KMS\ were obtained from near- and mid-IR lines (Jennings et al. 
\cite{jennings86}; Jennings \& Sada \cite{jennings98}; 
Ryde et al. \cite{ryde06a}; Tsuji \cite{tsuji06}).

Josselin \& Plez (\cite{josselin07}) analyzed spectral line profiles in the 
optical and extracted information about the velocity field in a sample of 
RSGs.  
For Betelgeuse, they found that the strong lines 
with lower excitation potentials of $\sim$1~eV show  blue- and red-shifted 
components at approximately $\pm 10$~\KMS\ with respect to the velocity 
of weak lines with excitation potentials of $\sim$3~eV.  
These velocities 
roughly agree with the 10--15~\KMS\ derived from our modeling 
of the CO first overtone lines with excitation potentials of $<$1~eV.  
Josselin \& Plez (\cite{josselin07}) detected no periodic or regular 
temporal variation in the velocity of the blue- and red-shifted components, 
which led them to interpret these two components as upward 
and downward convective motion.  

On the other hand, 
it is not so obvious whether or not such strong convective motion 
can be present in the CO line formation layers, where the convective 
energy flux is small.  
Alternatively, it is possible that the CO gas motion detected by our AMBER 
observations may represent local mass ejections.  
High-resolution studies of the dusty RSG VY~CMa and less massive 
evolved stars such as IRC+10216 and CIT6 
suggest that the mass loss is accompanied by episodic mass 
ejections in clumps or arcs (Humphreys et al. \cite{humphreys07}; 
Smith et al \cite{smith09}; Weigelt et al. \cite{weigelt02}; 
Monnier et al. \cite{monnier00}).  
While the circumstellar envelope around Betelgeuse is spherical on a 
large scale unlike VY~CMa, inhomogeneities such as clumps and plumes have 
been detected (Plez \& Lambert \cite{plez02}; Smith et al. \cite{smith09}).  
The long-term variability of the H$\alpha$ line velocity can also be 
interpreted as a consequence of ``intermittent failed ejections'', 
in which material is flung out and falls back toward the star 
(Smith et al. \cite{smith89}).  

To summarize, our AMBER observations are the first 
spatially-resolved detection of macroturbulent gas motion 
in a stellar atmosphere (photosphere and possibly MOLsphere) other than the 
Sun\footnote{The chromospheric gas motion toward Betelgeuse was spatially 
resolved with the HST observations in the UV by 
Gilliland \& Dupree (\cite{gilliland96}).}. 
The spatially resolved CO gas motion can represent the convective motion in 
the photosphere (and also in the MOLsphere) or the motion related to 
intermittent mass ejections in clumps and/or arcs.

The $(u,v)$ coverage of the present AMBER data is 
insufficient to derive the actual inhomogeneous structure in the photosphere 
and MOLsphere.  
In particular, the lack of visibilities in the 
first lobe makes it difficult to measure the size of the MOLsphere seen 
in the CO lines, and we simply assumed the same radius and temperature 
as those derived from the near- and mid-IR \HOH\ 
features.  While baselines shorter than the 16~m currently available at the 
VLTI is desirable, it is also possible to observe Betelgeuse in the 
first lobe by taking advantage of the projection effect.  Such new data 
will be indispensable for constraining the geometrical extent of the CO gas 
in the MOLsphere of Betelgeuse.

Lastly, we estimate the number of observations necessary to reconstruct 
an image.  
As a rule of thumb, the number of $(u,v$) points should be larger than the 
number of filled pixels (i.e., pixels with stellar flux) in the reconstructed 
image (e.g., Haniff \cite{haniff07}).  
This means that we need at least five $(u,v)$ points---realistically about 
10 points---to reconstruct a 1-D image
with five resolution elements over the stellar disk 
as in our present work.  
Therefore, to reconstruct a 2-D image with 5$\times$5 pixels, we need 
approximately 100 $(u,v)$ points, which corresponds to $\sim$33 observations 
($\sim$3 nights) 
with AMBER using three telescopes.  
It is crucial that these $(u,v)$ points are as uniformly distributed 
as possible, which will become feasible when more new VLTI 
configurations, particularly 
short baselines perpendicular to E0-G0-H0, are opened.  
The above estimate is roughly consistent with 
the image reconstruction simulation for the VLTI 2nd generation instrument 
MATISSE (Hofmann et al. \cite{hofmann08}).

\section{Concluding remarks}

We have spatially resolved the CO gas motion in the atmosphere of Betelgeuse 
with high-spectral resolution using 
VLTI/AMBER and successfully measured the visibilities, DPs, 
and CPs in the second, third, and fifth lobes, 
marking the highest spatial resolution ($\sim$9~mas) achieved for 
Betelgeuse.  
The visibilities observed in the CO first overtone lines 
suggest that the blue and red wings of individual lines originate in 
spatially distinct regions, and the non-zero/non-$\pi$ DPs and CPs 
observed in the CO lines corroborate this picture.  
Our simple model suggests an inhomogeneous velocity field with amplitudes 
of $\sim$10--15~\KMS\ in the atmosphere of Betelgeuse.  
These AMBER observations are the first to spatially resolve the so-called 
macroturbulence in a stellar atmosphere other than the Sun.  
The spatially resolved CO gas motion is likely to correspond to the 
convective motion in the upper 
photosphere (and possibly MOLsphere as well) or intermittent, failed 
clumpy mass ejections. 
Our modeling also shows that the AMBER data are consistent with the 
presence of the MOLsphere extending to $\sim$1.45~\RSTAR\ 
with a CO column density of $\sim \! 10^{20}$~\PERSQCM\ and a temperature 
of 1800~K.  

The visibilities and CPs observed in 
the continuum below 2.293~\micron\ do not show a drastic deviation from 
a limb-darkened disk with a diameter of 43.56~mas and a limb-darkening 
parameter of 0.12 at the time of the observations.  
Comparison of our AMBER HR\_K data with the previous AMBER 
MR\_K data, VINCI data, and IOTA measurements reveals that 
the recent 3-D convection simulations for Betelgeuse predict 
the $K$-band continuum visibility to be too low beyond the first null.

We plan to continue AMBER observations of Betelgeuse to study 
temporal variations in the visibilities and phases in the CO lines 
as well as in the continuum.  Such high-spatial resolution data 
will provide tight constraints on the time scale and spatial scale 
of the inhomogeneities.  
When more baselines, particularly short ones, become available, 
reconstruction of a $5 \times 5$-pixel image will be feasible 
with observations in $\sim$3~nights.  High-resolution imaging 
for various molecular lines will be essential for understanding 
the physical processes responsible for the inhomogeneous structures 
in RSGs.

\begin{acknowledgement}

We thank the ESO VLTI team in Garching and in Paranal, particularly 
F.~Rantakyr\"o, for supporting our AMBER observations.  
We also thank Jeff Meisner for his reduction of the VINCI 
data and the discussion about the accuracy of the determination of 
the diameter diameters from the VINCI data. 

\end{acknowledgement}

\clearpage

\clearpage

\clearpage

\clearpage
\begin{figure*}
\resizebox{\hsize}{!}{\rotatebox{-90}{\includegraphics{12247F5.ps}}}
\caption{Comparison between our patchy model (\TIN\ = 2250~K, \NCOOUT\ = 
$1 \times 10^{20}$~\PERSQCM, \VFLOW\ = 10~\KMS, $\Theta$ = 60\degr, $\theta$ = 
40\degr, and $\phi$ = 10\degr) and the AMBER data for 
Betelgeuse.  In all panels, the solid lines represent the model, while 
the dots represent the observational data (data set \#1).  
{\bf a:} Normalized flux.  
{\bf b--d:} Visibilities on the E0-G0-16m, G0-H0-32m, and 
E0-H0-48m baselines.  The observed and model visibilities on the latter two 
baselines are binned with three and five pixels, respectively. 
{\bf e:} Closures phase.  The observed data and the model are binned 
with five pixels.  
{\bf f--h:} Differential phases on the E0-G0-16m, G0-H0-32m, and 
E0-H0-48m baselines.  The observed and model DPs on the latter two baselines 
are binned with three and five pixels, respectively.  
}
\label{alfori_model}
\end{figure*}

\clearpage
\begin{figure}
\resizebox{\hsize}{!}{\rotatebox{0}{\includegraphics{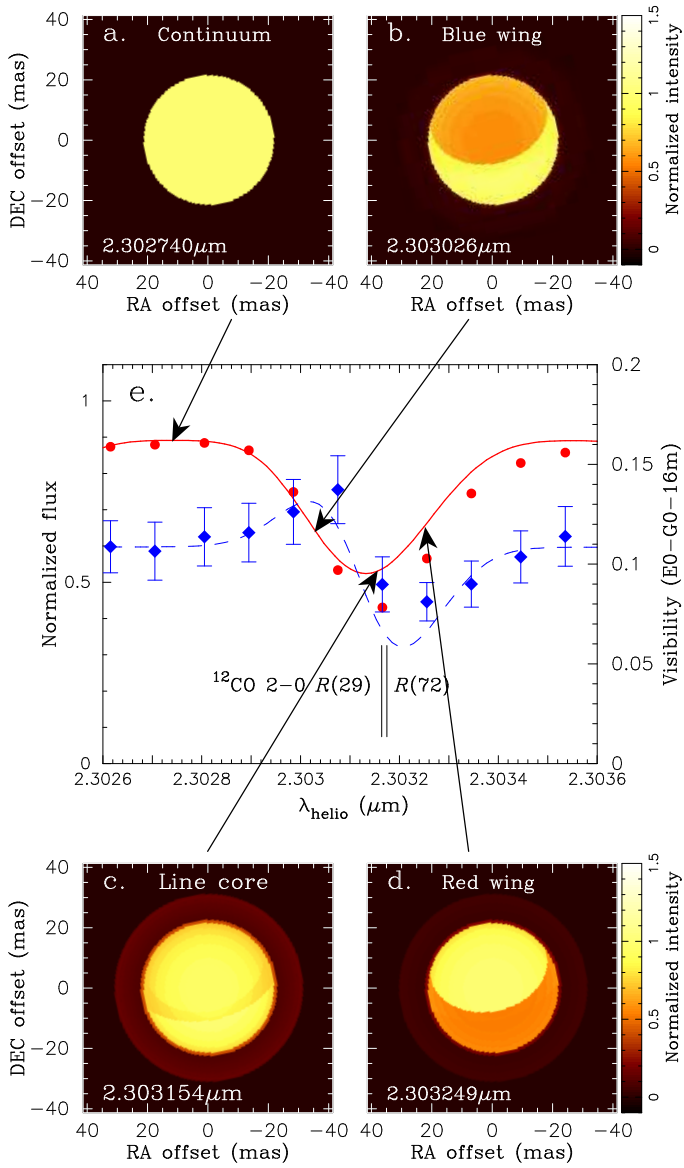}}}
\caption{Images of Betelgeuse predicted by our patchy model shown in 
Fig.~\ref{alfori_model} at different 
wavelengths in the CO line consisting of $R$(29)  and $R$(72).  
{\bf a--d:} Model images in the continuum, blue wing, line core, and 
red wing as indicated in the panel {\bf e}. 
The extended MOLsphere is slightly visible in the panel {\bf c}.  
{\bf e:} Normalized flux (left ordinate) and visibility on the shortest 
baseline (right ordinate).  
The filled circles and solid line represent the observed and model flux, 
respectively.  
The observed visibility and model prediction are shown by the filled 
diamonds and dashed line, respectively.  
The line positions are marked with the vertical ticks.  
A color version of this figure is available in the electronic edition. 
}
\label{alfori_images}
\end{figure}

\clearpage
\begin{table*}
\begin{center}
\caption {Summary of the AMBER observations of Betelgeuse and the calibrator 
Sirius.  
$t_{\rm obs}$: time of observation (Coordinated Universal Time=UTC), 
Tel.: telescope configuration, 
$B_{\rm p}$: projected baseline length, 
P.A.: position angle of the projected baseline on the sky, 
seeing in the visible, 
DIT: Detector Integration Time, 
Frames: the number of frames.  
}
\vspace*{-2mm}

\begin{tabular}{l l c c c c l l l r r}\hline
\# & Name & Night  & $t_{\rm obs}$ & Tel. & $B_{\rm p}$ & P.A.   & Seeing &
Airmass & DIT & Frames\\ 
   &      &        & (UTC)         &      & (m)         & (\degr)&  (\arcsec)  & &
   (ms) & \\
\hline
\multicolumn{11}{c}{2008, \HRK\ ($\lambda /\Delta \lambda$ = 12000), 
2.28--2.31~\micron}\\
\hline
1  & Betelgeuse & Jan. 08 & 03:52:58 & E0-G0-H0 & 16.00/31.97/47.96 & 73
& 0.4 & 1.19 & 120 & 2500\\
Cal  & Sirius &  Jan. 08 & 04:18:54 & E0-G0-H0 & 16.00/31.99/47.99 & 72 
& 0.3 & 1.01 & 120 & 2500\\
2  & Betelgeuse & Jan. 08 & 04:48:24 & E0-G0-H0 & 15.70/31.37/47.06 & 71 
& 0.3 & 1.28 & 120 & 2500\\
\hline
\multicolumn{11}{c}{2006, \MRK\ ($\lambda /\Delta \lambda$ = 1500), 
2.1--2.2~\micron}\\
\hline
3  & Betelgeuse & Feb. 10 & 02:45:08 & E0-G0    & 15.45             & 70 
& 1.1 & 1.33 & 48 & 6000 \\
Cal  & Sirius     & Feb. 10 & 04:50:22 & E0-G0    & 12.73             & 81 
& 1.1 & 1.32 & 48 & 2500 \\
\hline
\label{obslog}
\vspace*{-7mm}

\end{tabular}
\end{center}
\end{table*}

\clearpage
\begin{table}
\begin{center}
\caption {Summary of the VINCI observations of Betelgeuse. 
$t_{\rm obs}$: time of observation, 
Stations: siderostat stations, 
$B_{\rm p}$: projected baseline length, 
P.A.: position angle of the projected baseline on the sky, and 
Visibilities: the result based on the wavelet analysis. 
}
\vspace*{-2mm}

\begin{tabular}{r l l r r l }\hline
\# & $t_{\rm obs}$ & Stations & $B_{\rm p}$ & P.A. & Visibility \\
   & (UTC)        &       & (m)       & (\degr) &          \\
\hline
\multicolumn{6}{c}{2001 Oct. 12}\\
\hline
 1 &   08:11:44 & E0-G0  &  14.71 &   74 & $  0.0988\pm  0.0021$ \\
\hline
\multicolumn{6}{c}{2002 Dec. 06}\\
\hline
 2 &   05:10:46 & B3-C3  &   7.70 &   74 & $  0.4875\pm  0.0066$ \\
 3 &   05:37:38 & B3-C3  &   7.89 &   73 & $  0.4637\pm  0.0086$ \\
\hline
\multicolumn{6}{c}{2003 Jan. 07}\\
\hline
 4 &   02:40:04 & B3-C3  &   7.50 &   74 & $  0.4752\pm  0.0082$ \\
\hline
\multicolumn{6}{c}{2003 Jan. 09}\\
\hline
 5 &   01:41:42 & B3-C3  &   6.79 &   74 & $  0.5653\pm  0.0069$ \\
 6 &   02:10:29 & B3-C3  &   7.25 &   74 & $  0.5160\pm  0.0063$ \\
\hline
\multicolumn{6}{c}{2003 Jan. 13}\\
\hline
 7 &   02:02:03 & B3-C3  &   7.25 &   74 & $  0.5128\pm  0.0064$ \\
 8 &   02:05:11 & B3-C3  &   7.32 &   74 & $  0.5093\pm  0.0063$ \\
 9 &   02:11:44 & B3-C3  &   7.39 &   74 & $  0.5060\pm  0.0063$ \\
10 &   02:15:32 & B3-C3  &   7.44 &   74 & $  0.5060\pm  0.0063$ \\
11 &   02:19:20 & B3-C3  &   7.48 &   74 & $  0.4989\pm  0.0062$ \\
12 &   03:23:52 & B3-C3  &   7.96 &   73 & $  0.4503\pm  0.0056$ \\
13 &   03:27:39 & B3-C3  &   7.97 &   73 & $  0.4461\pm  0.0055$ \\
14 &   03:31:28 & B3-C3  &   7.97 &   72 & $  0.4463\pm  0.0055$ \\
15 &   04:00:31 & B3-C3  &   7.97 &   72 & $  0.4546\pm  0.0057$ \\
16 &   04:04:18 & B3-C3  &   7.96 &   71 & $  0.4519\pm  0.0057$ \\
17 &   04:08:07 & B3-C3  &   7.95 &   71 & $  0.4476\pm  0.0056$ \\
18 &   04:33:46 & B3-C3  &   7.83 &   70 & $  0.4614\pm  0.0058$ \\
19 &   04:35:47 & B3-C3  &   7.81 &   70 & $  0.4666\pm  0.0058$ \\
20 &   04:39:36 & B3-C3  &   7.78 &   70 & $  0.4719\pm  0.0059$ \\
\hline
\multicolumn{6}{c}{2003 Jan. 14}\\
\hline
21 &   00:49:57 & B3-C3  &   6.04 &   73 & $  0.6780\pm  0.0068$ \\
22 &   00:54:02 & B3-C3  &   6.13 &   73 & $  0.6691\pm  0.0068$ \\
23 &   00:58:04 & B3-C3  &   6.21 &   73 & $  0.6560\pm  0.0067$ \\
24 &   01:09:58 & B3-C3  &   6.46 &   73 & $  0.5993\pm  0.0055$ \\
25 &   01:14:00 & B3-C3  &   6.53 &   74 & $  0.5942\pm  0.0054$ \\
26 &   01:18:01 & B3-C3  &   6.61 &   74 & $  0.5870\pm  0.0054$ \\
27 &   01:27:23 & B3-C3  &   6.78 &   74 & $  0.5779\pm  0.0055$ \\
28 &   01:31:26 & B3-C3  &   6.85 &   74 & $  0.5802\pm  0.0057$ \\
29 &   01:41:48 & B3-C3  &   7.02 &   74 & $  0.5478\pm  0.0051$ \\
30 &   01:45:52 & B3-C3  &   7.08 &   74 & $  0.5432\pm  0.0051$ \\
31 &   01:49:54 & B3-C3  &   7.14 &   74 & $  0.5350\pm  0.0052$ \\
32 &   02:54:37 & B3-C3  &   7.84 &   73 & $  0.4871\pm  0.0065$ \\
33 &   02:58:40 & B3-C3  &   7.86 &   73 & $  0.4811\pm  0.0059$ \\
34 &   03:02:44 & B3-C3  &   7.89 &   73 & $  0.4940\pm  0.0057$ \\
35 &   03:14:34 & B3-C3  &   7.94 &   73 & $  0.4534\pm  0.0043$ \\
36 &   03:18:37 & B3-C3  &   7.95 &   73 & $  0.4534\pm  0.0043$ \\
37 &   03:22:39 & B3-C3  &   7.96 &   73 & $  0.4573\pm  0.0044$ \\
38 &   03:34:14 & B3-C3  &   7.98 &   72 & $  0.4649\pm  0.0045$ \\
39 &   03:38:18 & B3-C3  &   7.99 &   72 & $  0.4582\pm  0.0043$ \\
40 &   03:42:22 & B3-C3  &   7.99 &   72 & $  0.4550\pm  0.0044$ \\
41 &   04:20:35 & B3-C3  &   7.88 &   71 & $  0.4530\pm  0.0044$ \\
42 &   04:24:38 & B3-C3  &   7.86 &   70 & $  0.4643\pm  0.0043$ \\
43 &   04:28:43 & B3-C3  &   7.83 &   70 & $  0.4688\pm  0.0046$ \\
\hline
\label{obslog_vinci}
\vspace*{-7mm}

\end{tabular}
\end{center}
\end{table}
\addtocounter{table}{-1}

\begin{table}
\begin{center}
\caption {Continued}
\vspace*{-2mm}
\begin{tabular}{r l l r r l }\hline
\# & $t_{\rm obs}$ & Stations & $B_{\rm p}$ & P.A. & Visibility \\
   & (UTC)        &       & (m)       & (\degr) &          \\
\hline
\multicolumn{6}{c}{2003 Jan. 15}\\
\hline
44 &   00:52:16 & B3-C3  &   6.17 &   73 & $  0.6375\pm  0.0087$ \\
45 &   00:56:20 & B3-C3  &   6.26 &   73 & $  0.6444\pm  0.0088$ \\
46 &   01:00:21 & B3-C3  &   6.34 &   73 & $  0.6269\pm  0.0086$ \\
47 &   01:14:12 & B3-C3  &   6.61 &   74 & $  0.6642\pm  0.0109$ \\
48 &   01:18:14 & B3-C3  &   6.68 &   74 & $  0.5902\pm  0.0080$ \\
49 &   01:31:23 & B3-C3  &   6.92 &   74 & $  0.5704\pm  0.0075$ \\
50 &   01:35:30 & B3-C3  &   6.98 &   74 & $  0.5612\pm  0.0074$ \\
51 &   01:45:14 & B3-C3  &   7.15 &   74 & $  0.5252\pm  0.0069$ \\
52 &   01:50:55 & B3-C3  &   7.22 &   74 & $  0.5228\pm  0.0068$ \\
53 &   01:56:38 & B3-C3  &   7.30 &   74 & $  0.5110\pm  0.0067$ \\
54 &   02:07:02 & B3-C3  &   7.43 &   74 & $  0.4975\pm  0.0065$ \\
55 &   02:12:43 & B3-C3  &   7.51 &   74 & $  0.4847\pm  0.0063$ \\
56 &   02:18:24 & B3-C3  &   7.57 &   74 & $  0.4898\pm  0.0064$ \\
57 &   03:57:19 & B3-C3  &   7.96 &   71 & $  0.4525\pm  0.0059$ \\
58 &   04:03:01 & B3-C3  &   7.94 &   71 & $  0.4555\pm  0.0059$ \\
59 &   04:08:42 & B3-C3  &   7.91 &   71 & $  0.4618\pm  0.0060$ \\
60 &   04:19:11 & B3-C3  &   7.87 &   70 & $  0.4578\pm  0.0060$ \\
61 &   04:25:17 & B3-C3  &   7.83 &   70 & $  0.4765\pm  0.0063$ \\
62 &   04:30:58 & B3-C3  &   7.78 &   70 & $  0.4876\pm  0.0066$ \\
\hline
\multicolumn{6}{c}{2003 Jan. 18}\\
\hline
63 &   01:11:11 & B3-C3  &   6.76 &   74 & $  0.5671\pm  0.0106$ \\
64 &   01:14:20 & B3-C3  &   6.86 &   74 & $  0.5625\pm  0.0104$ \\
65 &   01:28:58 & B3-C3  &   7.14 &   74 & $  0.5296\pm  0.0097$ \\
66 &   01:48:44 & B3-C3  &   7.41 &   74 & $  0.5002\pm  0.0092$ \\
67 &   02:03:30 & B3-C3  &   7.60 &   74 & $  0.4676\pm  0.0086$ \\
68 &   03:13:00 & B3-C3  &   7.97 &   72 & $  0.4430\pm  0.0081$ \\
69 &   03:15:42 & B3-C3  &   7.98 &   72 & $  0.4389\pm  0.0081$ \\
70 &   03:52:06 & B3-C3  &   7.94 &   71 & $  0.4433\pm  0.0082$ \\
71 &   03:55:13 & B3-C3  &   7.93 &   71 & $  0.4458\pm  0.0083$ \\
72 &   04:00:32 & B3-C3  &   7.89 &   71 & $  0.4513\pm  0.0082$ \\
73 &   04:12:01 & B3-C3  &   7.81 &   70 & $  0.4557\pm  0.0083$ \\
\hline
\label{obslog_vinci2}
\vspace*{-7mm}

\end{tabular}
\end{center}
\end{table}

\begin{table}
\begin{center}
\caption {Calibrators used for the VINCI observations of Betelgeuse.  
Name of the calibrators, spectral type (Sp.Type), uniform-disk 
diameters ($d_{\rm UD}$), which are taken from Richichi \& Percheron 
(\cite{richichi05}).  
The number of the observations of the individual calibrators on each night 
is given in braces.  
}
\vspace*{-2mm}
\begin{tabular}{r l l l}\hline
Name & Sp.Type & $d_{\rm UD}$ (mas) & Night \\
\hline
$\varepsilon$ Peg & K2Ib & $7.70 \pm 0.24$     & 2001 Oct. 12 (4)\\
$\beta$ Cet       & G9II-III & $5.18 \pm 0.05$ & 2001 Oct. 12 (6)\\
                  &          &                 & 2002 Dec. 06 (2)\\
$\varepsilon$ Lep & K4III & $5.90 \pm    0.06$ & 2001 Oct. 12 (3)\\
                  &       &                    & 2002 Dec. 06 (1)\\
                  &       &                    & 2003 Jan. 09 (1)\\
Sirius            & A1V   & $5.60 \pm 0.15$    & 2001 Oct. 12 (4)\\
                  &       &                    & 2002 Dec. 06 (1)\\
                  &       &                    & 2003 Jan. 07 (1)\\
                  &       &                    & 2003 Jan. 18 (6)\\
$\delta$ CMa      & F8Iab & $3.29 \pm  0.46 $  & 2001 Oct. 12 (2)\\
                  &       &                    & 2003 Jan. 07 (1)\\

$\pi$ Eri         & M1III & $4.8  \pm   0.5 $  & 2002 Dec. 06 (1)\\
$\tau$ Pup        & K1III & $4.38 \pm   0.07$  & 2002 Dec. 06 (1)\\
                  &       &                    & 2003 Jan. 09 (1)\\
Procyon           & F5IV-V & $5.37 \pm 0.11 $  & 2002 Dec. 06 (1)\\
                  &       &                    & 2003 Jan. 07 (1)\\
1 Pup             & K5III & $3.8  \pm   0.4 $  & 2002 Dec. 06 (1)\\
$\chi$ Phe        & K5III & $2.69 \pm   0.03 $ & 2002 Dec. 06 (1)\\
$\iota$ Cet       & K1.5III & $3.28 \pm  0.04$ & 2003 Jan. 07 (1)\\
31 Ori            & K5III & $3.55 \pm 0.06$    & 2003 Jan. 07 (1)\\
$\theta$ CMa      & K4III & $4.13 \pm   0.4$   & 2003 Jan. 07 (1)\\
$\beta$ Cnc       & K4III & $4.88 \pm   0.03$  & 2003 Jan. 07 (1)\\
                  &       &                    & 2003 Jan. 09 (2)\\
$\zeta$ Hya       & G9II-III & $3.1 \pm   0.2$ & 2003 Jan. 07 (1)\\
                  &       &                    & 2003 Jan. 09 (1)\\
$\alpha$ Hya      & K3II-III & $9.44 \pm   0.9$& 2003 Jan. 07 (1)\\
                  &       &                    & 2003 Jan. 09 (6)\\
                  &       &                    & 2003 Jan. 13 (6)\\
$\iota$ Hya       & K2.5III & $3.41 \pm   0.05$& 2003 Jan. 07 (2)\\
                  &       &                    & 2003 Jan. 09 (4)\\
                  &       &                    & 2003 Jan. 13 (4)\\
$\mu$ Hya         & K2.5III & $4.69 \pm   0.5$ & 2003 Jan. 09 (3)\\
                  &       &                    & 2003 Jan. 13 (9)\\
                  &       &                    & 2003 Jan. 15 (2)\\
$\gamma^1$ Leo    & K1IIIb & $7.70 \pm 0.70$   & 2003 Jan. 09 (3)\\
$\beta$ Ori       & B8Iab  & $2.43 \pm   0.05$ & 2003 Jan. 13 (16)\\
                  &       &                    & 2003 Jan. 14 (22)\\
                  &       &                    & 2003 Jan. 15 (15)\\
V337 Car          & K3IIa & $5.09 \pm   0.06$  & 2003 Jan. 14 (6)\\
N Vel             & K5III & $6.92 \pm 0.08$    & 2003 Jan. 07 (1)\\
\hline
\label{cal_vinci}
\vspace*{-7mm}

\end{tabular}
\end{center}
\end{table}

\end{document}